\documentclass[aps,showpacs,superscriptaddress,prb,twocolumn]{revtex4-1}

\usepackage{amsmath}
\usepackage{amssymb}
\usepackage{graphicx}
\usepackage{epsfig}
\usepackage{dcolumn}
\usepackage{textcomp}
\usepackage{bm}
\usepackage[normalem]{ulem} 
\usepackage{color} 
\usepackage{subfigure}
\usepackage{pstricks}

\DeclareMathOperator{\sgn}{sgn}

\begin{document}

\title{The parity effect in Josephson junction arrays} 

\author{Jared H. Cole}
\email{jared.cole@rmit.edu.au}
\affiliation{%
Chemical and Quantum Physics, School of Applied Sciences, RMIT University, Melbourne, Victoria 3001, Australia.
}
\author{Andreas Heimes}
\affiliation{Institute f\"ur Theoretische Festk\"orperphysik, Karlsruhe Institute of Technology, 76128 Karlsruhe, Germany}
\author{Timothy Duty}
\affiliation{ARC Centre of Excellence for Engineered Quantum Systems and School of Physics,
University of New South Wales, Sydney, New South Wales 2052, Australia}
\author{Michael Marthaler}
\affiliation{Institute f\"ur Theoretische Festk\"orperphysik, Karlsruhe Institute of Technology, 76128 Karlsruhe, Germany}



\date{\today}
             
\begin{abstract}
We study the parity effect and transport due to quasiparticles in circuits comprised of many superconducting islands. We develop a general approach and show that it is equivalent to previous methods for describing the parity effect in their more limited regimes of validity.  As an example we study transport through linear arrays of Josephson junctions in the limit of negligible Josephson energy and observe the emergence of the parity effect with decreasing number of non-equilibrium quasiparticles.  Due to the exponential increase in the number of relevant charge states with increasing length, in multi-junction arrays the parity effect manifests in qualitatively different ways to the two junction case.  The role of charge disorder is also studied as this hides much of the parity physics which would otherwise be observed.  Nonetheless, we see that the current through a multi-junction array at low bias is limited by the formation of meta-stable even-parity states.
\end{abstract}

\maketitle

In superconducting circuits of small dimensions, charging effects play an important role. On the one hand the Coulomb blockade leads to charge pinning and an effective suppression of electronic transport at low bias voltage. On the other hand the superconducting nature manifests in the parity effect, i.e. given a odd number of electrons on the superconductor there is one remaining quasiparticle dominating the transport properties in the low bias regime~\cite{Averin:1994wh,Matveev:1993vr,Amar:1994wh,Glazman:1994tj,Tuominen:1993tc,Tuominen:1992vy,Lu:1998tj,Lu:1996uu}. Strictly speaking this picture is true for equilibrium and very low temperatures. However if a non-equilibrium situation is imposed, e.g. by applying a finite bias voltage, the average number of quasiparticles may be increased. Recently such non-equilibrium quasiparticle effects have been investigated in superconducting qubits~\cite{Lutchyn:2007gw,Lutchyn:2007fc,Shaw:2008il,Catelani:2011tl,Catelani:2011th,Ansari:2015cj} and single electron transistors (SETs)~\cite{Lutchyn:2007ik,Court:2008vi,Manninen:2012ti,Maisi:2013vu,Heimes:2014fc}. 
In this context the interplay of charge transport, the excitation of non-equilibrium quasiparticles and the observation of the parity effect has been the subject of recent experiments with SETs~\cite{Maisi:2013vu}. Based on related theoretical modelling~\cite{Janko:1994uy, Golubev:1994dy,Heimes:2014fc} we extend the prevailing transport theory of multi-junction circuits and show that this approach removes the ambiguities of previous approaches when including parity effects for more than one superconducting island, although our approach is equivalent to earlier work in the appropriate limits.  As an example, we perform the first analysis of the parity effect in linear multi-junction arrays and make a number of predictions for the electronic transport signatures that can be identified with the parity effect in these systems.   


\begin{figure} [t!]
\centering{\includegraphics[width=1.0\columnwidth]{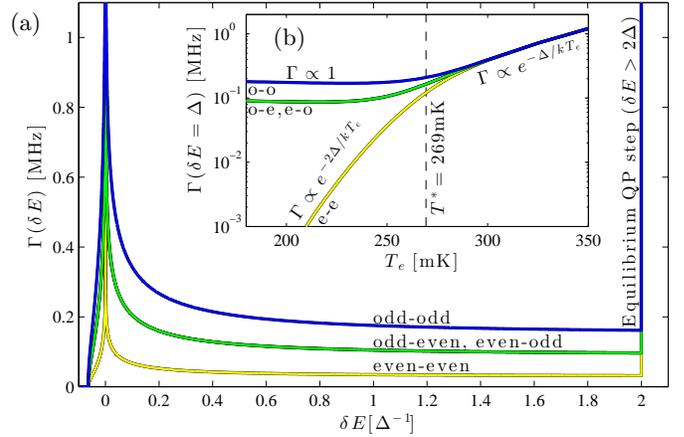}}
\rput[tl](-4.5,6){(a)}
\rput[tl](-1,5.8){(b)}
\caption{(a) Quasiparticle tunnelling rate (Eq.~\ref{eq:Rate}) as a function of energy difference between charge configurations ($\delta E$) for four different parity configurations, ie.\ odd or even charge states on the source and destination islands respectively.  The rates are evaluated at $T_e=222$ mK which is just below the parity temperature $T^{*}=269$ mK for these parameters - see text for details. (b) Parity rates evaluated in the middle of the sub-gap region ($\delta E = \Delta$) as a function of $T_e$.  The scaling behaviour for the four rates as a function of $\Delta/k_B T_e$ can be clearly seen above and below the parity temperature. \label{fig:rates}}
\end{figure}


\section{The parity effect in multi-junction circuits.}
Conventional equilibrium quasiparticle theory states that transport through a Josephson junction is exponentially suppressed when the applied bias $V$ across the junction is less than twice the superconducting gap, $\Delta$.  However upon closer inspection there is a measurable sub-gap transport contribution for $V<2\Delta$, which depends strongly on the charge state parity of the islands either side of the junction.
This parity effect is traditionally modelled in single-electron transistors using one of two approaches.  

In the first approach, the free energy of the circuit gains an additional free energy contribution (which depends on temperature, superconducting gap and island volume) due to the parity of the occupancy of the island~\cite{Averin:1992uo,Hekking:1993wi,Lafarge:1993jj,Matveev:1993vr,Glazman:1994tj,Lu:1996uu,Tinkham:1995wa,Feigelman:1997ta,Lu:1998tj,Lambert:2014gg}.  In the alternative approach\cite{Schon:1994th,Schon:1994tr,Lutchyn:2007ik,Lutchyn:2007gw}, the rates associated with both the equilibrium quasiparticles and the odd quasiparticle state `at gap' must be computed separately taking into account the relative chemical potential differences between island and lead.  Although in SETs both approaches describe similar physics; in the multi-junction case significant complications (both conceptual and technical) arise when applying either of these methods.  In this work we show how to describe the parity contribution in a general way, as well as show how these earlier methods are considered limiting cases of the theory as presented here.

Following more recent work on normal-superconducting-normal SETs\cite{Maisi:2013vu}, sub-gap quasiparticle effects can be included in a consistent way such that the rate for an arbitrary charge transfer event is computed based on the initial parity of the origin and destination islands.  Throughout this discussion, we parameterize the distribution of non-equilibrium quasiparticles by an effective electron temperature $T_e$, rather than explicitly keeping track of the non-equilibrium quasiparticle number on each island as was considered in Ref.~\onlinecite{Heimes:2014fc}.  This formalism is also applicable when modelling the parity corrections to sub-gap quasiparticle transport at base temperatures beyond the regime typically associated with the parity effect itself.


To include the contribution from parity dependent quasiparticle tunnelling in a general way, we scale the Fermi function for each island in the circuit by a factor $A_j$, which depends on whether the charge state of the island $n$ is odd or even ($j=n \!\mod 2$).  This accounts for the fact that in the odd charging state there is at least one quasiparticle remaining unpaired (see Ref.~\onlinecite{Heimes:2014fc} for further details).  This modification of the Fermi function $f(E) \rightarrow A_j f(E)$, is strictly true only for $E>0$ and we define,
\begin{equation}
A_j = \left[ \tanh(N_{qp}) \right]^{(-1)^j}
\end{equation}  
where we parameterise the average number of excited quasiparticles $N_{qp}$ as~\cite{Maisi:2013vu,Heimes:2014fc}
\begin{equation}\label{eq:Nqp}
N_{qp} = N(0) V \sqrt{2 \pi \Delta k_{B} T_e} \exp\left[ -\frac{\Delta}{k_B T_e} \right].
\end{equation}
Here, the average number of excited quasiparticles is expressed in terms of the superconducting density of states evaluated at the Fermi level $N(0)$, the volume of the island $V$, the superconducting gap $\Delta$ and the effective temperature $T_e$ of the quasiparticles. When the number of excited quasiparticles (Eq.~\ref{eq:Nqp}) is less than one, the scaling factor $A_j$ shows markedly different behaviour for even and odd charging states, leading to parity dependent transport signatures.  The crossover temperature $T^{*}$ below which these effects can be observed is given by,
\begin{equation}
T^{*} \approx \frac{\Delta}{k_{B} \ln(N(0) V \sqrt{2 \pi \Delta k_{B} T^{*}})}
\end{equation} 
which must be solved self-consistently.  In the limit of electron temperature $T_e > T^{*}$, the even-odd distinction vanishes and therefore the parity effect is unobservable.  

Although expressing the non-equilibrium contribution in terms of $A_j$ is a very general approach, for $f(E>0)\ll1$ it proves to be both conceptually and computationally useful to parameterise the non-equilibrium quasiparticle distribution in terms of a modified chemical potential, $\mu_j$.  To do this we express $A_j$ in the form of a shifted Fermi distribution such that
\begin{equation}\label{eq:mu_j}
\mu_j(E) = \sgn(E) (-1)^j k_B T_e \ln \left[ \tanh(N_{qp}) \right] 
\end{equation}  
where the factor $(-1)^j$ takes into account the odd-even discrepancy and the $\sgn(E)$ term accounts for the fact that the $f(E) \rightarrow A_j f(E)$ replacement applies strictly to positive energy differences. 

To study parity effects in a general way which will be applicable to multi-junction circuits, the single-electron tunnelling rate between islands also possess a even/odd charge state dependence and in general is given by
\begin{widetext}
\begin{equation}\label{eq:Rate}
\Gamma_{n, m}(\delta E) = \frac{1}{e^2 R_T} \int_{-\infty}^{\infty} dE \frac{N(E)}{N(0)} \frac{N(E+\delta E)}{N(0)} f_e(E-\mu_n(E)) [1-f_e(E+\delta E-\mu_m(E+\delta E))] .
\end{equation}
\end{widetext}
This rate is expressed as a function of the energy difference $\delta E$ between initial ($n,m$) and final charge states ($n-1,m+1$ or $n+1,m-1$), where $f_e(E)$ is defined as the Fermi function at temperature $T_e$ and $R_T$ is the junction normal tunnel resistance.  In this context $n$ and $m$ indicate the \emph{initial} even-odd parity of the origin and destination islands respectively.

When considering the movement of a single charge between two islands, the rate given by Eq.~\ref{eq:Rate} depends on the initial parity of both islands - giving four possible rates.  If we consider the overall scaling of the sub-gap rates (insert to Fig.~\ref{fig:rates}), we see that above the parity temperature $T_e > T^{*}$, all four rates scale $\propto \exp[-\Delta/k_B T_e]$.  Below the parity temperature, the `even-even' rate scales $\propto \exp[-2 \Delta/k_B T_e]$, whereas the other three rates are approximately temperature independent for $T_e < T^{*}$.  

To understand this behaviour, we can approximate Eq.~\ref{eq:Rate} by expanding around the divergences in the BCS density of states, $N(E)/N(0)$.  Expanding $E=\Delta(1+\epsilon)$ and taking the dominant terms for each side of the density of states, we obtain,
\begin{equation}
\Gamma_{n,m}(\delta E) = \frac{\Delta}{e^2 R_T} \frac{N(\Delta + \delta E)}{N(0)} \int_0^\infty d\epsilon \frac{1}{\sqrt{\epsilon}} g(E, \delta E)
\end{equation}
where
\begin{widetext}
\begin{equation}
g(E, \delta E) =  f(-\Delta(1+\epsilon) - \delta E + \mu_n) [1 - f(-\Delta(1+\epsilon) + \mu_m)] + f(\Delta(1+\epsilon) - \mu_n) [1 - f(\Delta(1+\epsilon) + \delta E - \mu_m)].
\end{equation}  
\end{widetext}
We can then evaluate $g(E, \delta E)$ and therefore the integral in various limits of interest.

In the limit $T_e>T^{*}$, the parity dependent chemical potential term $\mu_n \approx 0$ and therefore $g(E, \delta E)$ simplifies considerably.  If we assume $\Delta \gg k_B T_e$ and $\delta E > \epsilon \Delta$, we obtain $g(E, \delta E) \approx \exp[-\epsilon \Delta / k_B T_e]$ and therefore 
\begin{equation}
\Gamma^{(T_e>T^{*})}_{n,m} \approx \frac{N(\Delta + \delta E)}{N(0)} \frac{N_{qp}}{e^2 R_T N(0) V}.
\end{equation}
The sub-gap quasiparticle rate as a function of $\delta E$ therefore takes on the functional form of the BCS density-of-states near the divergence, for temperatures above the parity temperature.  The magnitude of this rate scales proportional to the quasiparticle number $N_{qp}$ and as expected is independent of the parity of the source and destination islands.

Turning to the low temperature case ($T_e\ll T^{*}$), we now must deal with the various values of $\mu_{n,m}$.  
Taking the limit of $N_{qp} \ll 1$ we can express the chemical potential shift as
\begin{equation}\label{eq:mu_n}
\mu_n \approx (-1)^{n} [k_B T_e \ln (N_{\rm{eff}}) - \Delta]
\end{equation} 
where $N_{\rm{eff}}=N(0) V \sqrt{2 \pi \Delta k_{B} T_e}$.  Using this expression, we then evaluate $g(E, \delta E)$ for each of the four cases, in the limit that $\Delta \gg k_B T_e$, $\delta E > \epsilon \Delta$ and $N_{\rm{eff}}\gg1$.  This gives the following expressions,
\begin{equation}\label{Eq:eeRate}
\Gamma^{(T_e\ll T^{*})}_{e,e} \approx \frac{N(\Delta + \delta E)}{N(0)} \frac{N_{qp}^2}{e^2 R_T N(0) V}
\end{equation}
\begin{equation}\label{Eq:ooRate}
\Gamma^{(T_e\ll T^{*})}_{o,o} \approx \frac{N(\Delta + \delta E)}{N(0)} \frac{1}{e^2 R_T N(0)V}
\end{equation}
\begin{equation}\label{Eq:oeRate}
\Gamma^{(T_e\ll T^{*})}_{o,e} = \Gamma^{(T_e\ll T^{*})}_{e,o} = \frac{\Gamma^{(T_e\ll T^{*})}_{o,o} + \Gamma^{(T_e\ll T^{*})}_{e,e}}{2}.
\end{equation}
In all rates, we see the characteristic density of states dependence on $\delta E$ as well as a factor of $N_{qp}^2$ difference between the odd-odd and even-even rates.  Therefore the even-even rate scales with $N_{qp}^2 \propto \exp(-2 \Delta/k_B T_e)$ whereas the odd-odd rate is approximately constant approaching zero temperature.  As can be seen in Fig.~\ref{fig:rates}, the odd-even and even-odd rates are equal and $\approx \Gamma^{(T_e\ll T^{*})}_{o,o}/2$ (as the contribution from the even-even rate is negligible at low temperatures).

The approximate rates given above can be compared to previous work on the parity effect in simpler circuits.  The mapping to the free energy shift of $\Delta$ often ascribed to the odd charge state~\cite{Averin:1992uo,Hekking:1993wi,Lafarge:1993jj,Matveev:1993vr,Glazman:1994tj} follows directly from Eq.~\ref{eq:mu_n}.  However, we can immediately see the lack of generality of that approach because the sign of the shift depends on the parity of both the source and destination charge states. In our case of many islands a simple $\Delta$ correction for each odd charge state is manifestly not sufficient.

Previous work on applying a shifted chemical potential~\cite{Schon:1994th,Schon:1994tr, Siewert:1996tf} is in principle similar to our approach.  In that case the ``odd to even'' transition rate (meaning the transition from an odd to even state \emph{of the same island}) is equivalent to our Eq.~\ref{Eq:ooRate}.  However it is not clear how to easily generalise this method to multiple islands when the parity of both the source and destination islands must be taken into account.  Furthermore, Eqs.~\ref{Eq:eeRate}-\ref{Eq:oeRate} are only approximations to the general expression Eq.~\ref{eq:Rate}, due to the relatively crude approximation to the integral over $\epsilon$.  This becomes particularly important when comparing quantitatively to experiment.  


\section{The parity effect in Josephson junction arrays}
A Josephson junction array (JJA) is the multi-junction generalisation of the (superconducting) single-electron transistor.  Increasing the number of junctions changes the electrical response of such a circuit markedly~\cite{Likharev:1989tg,Bakhvalov:1989td,Delsing:1992tq,Haviland:1996wz} when compared to the simple two junction devices. New and interesting effects are observed, including hysteresis~\cite{Delsing:1992tq}, soliton propagation\cite{Haviland:1996wz,Hermon:1996up}, non-trivial magnetic field effects\cite{Haviland:1996wz, Bylander:2007gb,Shimada:2012tk} and correlated electron transport\cite{Bakhvalov:1989td,Bylander:2005gr,Bylander:2007gb}.  Detailed understanding of junction arrays also promise new superconducting devices; such as qubit designs based on large kinetic inductance~\cite{Manucharyan:2009fo,Manucharyan:cz,Ferguson:2013bd}, or terahertz radiation sources~\cite{Ozyuzer:2007jy,Wang:2010we,Yuan:2012dz}. Although the qualitative theory of Josephson junction arrays has been established for some time~\cite{Delsing:1992tq,Haviland:1996wz}, direct quantitative comparison between theory and experiment in these devices is still elusive and fraught with difficulty as they display qualitatively different physics than that seen in single-electron transistors (SETs) and other few junction devices.  

\begin{figure} [t!]
\rput[tl](0,2){(a)}
\rput[tl](0,0){(b)}
\centering{\includegraphics[width=1.0\columnwidth]{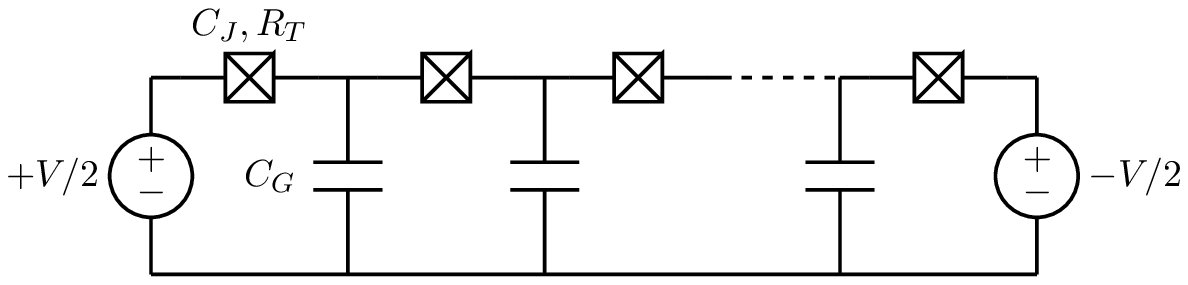}}\\
\vspace{0.3cm}
\centering{\includegraphics[width=1.0\columnwidth]{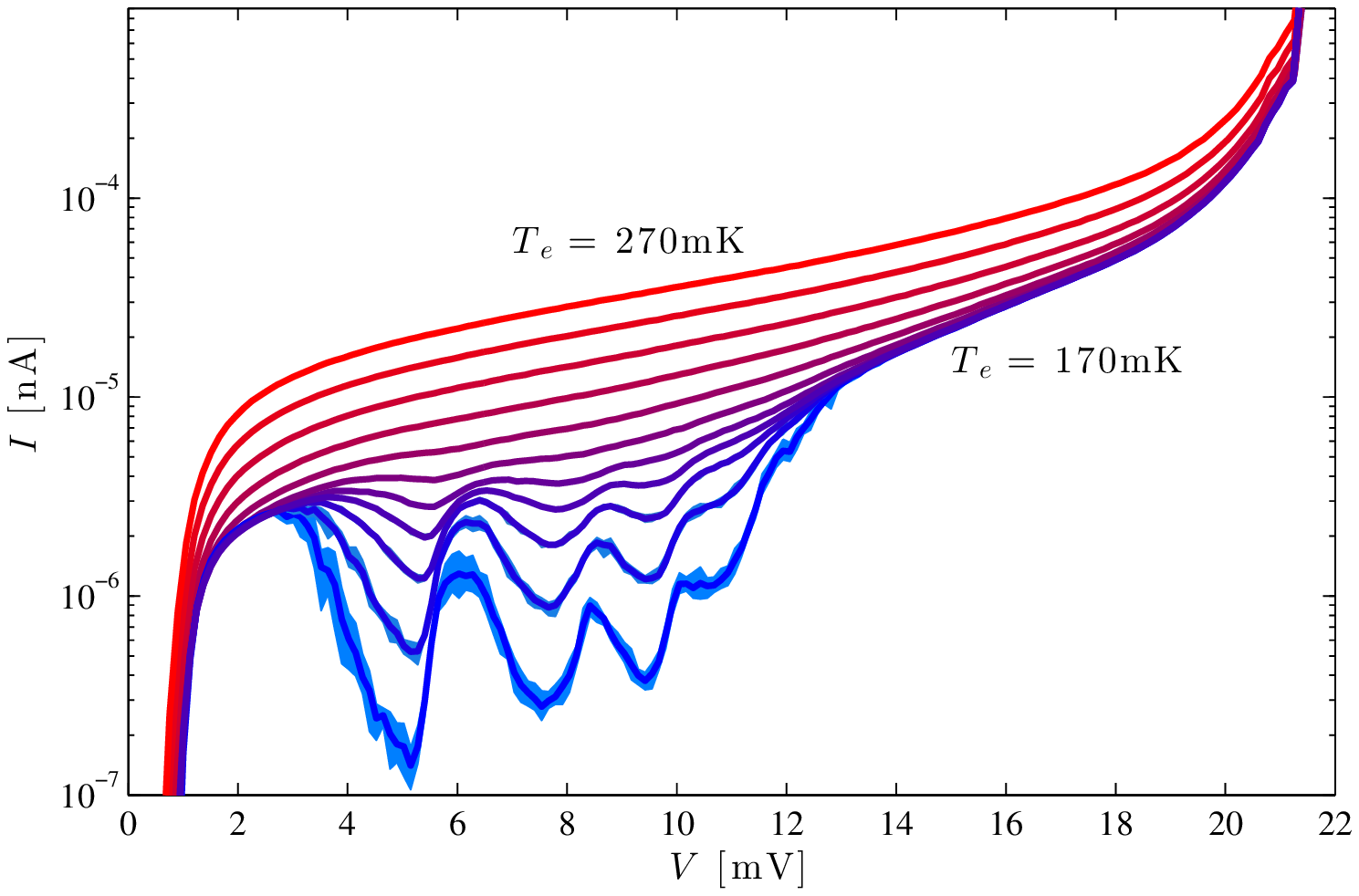}}
\caption{(a) JJA circuit under consideration, consisting of a linear chain of Josephson junctions with Josephson energy $E_J$ and capacitance $C_J$. The circuit is driven by a symmetrically applied voltage source $V$ and each junction sees an effective capacitance to ground $C_G$.  (b) Current through the array (logarithmic scale) as a function of applied voltage.  Effective electron temperature is varied from below to above the parity crossover temperature, which is $T^{*}=269$ mK for these parameters.  The lines in the figure are the average of 10 KMC runs, each of which consists of $10^6$ events.  The maximum and minimum of these 10 runs are indicated by the pale shading.\label{fig:paritywithnodisorder}
}
\end{figure}

To illustrate the manifestation of the parity effect in multi-island circuits, we simulate transport through a linear JJA, see Fig.\ref{fig:paritywithnodisorder}(a). We employ the kinetic Monte-Carlo method\cite{Bakhvalov:1989td,Wasshuber:2001ta,Mizugaki:2005vc,Voter:2007ul,Ho:2010ft,Reuter:2011vq,Cole:2014fk} following the procedure as detailed in Ref.\onlinecite{Cole:2014fk}.  The energy of various charge states of the array is computed based on purely electrostatic considerations as we confine our investigation to the negligible $E_J$ limit and therefore do not consider Cooper-pair transport (as $R_T=1$M$\Omega \gg R_Q = h/e^2$).  We assume a superconducting gap of $\Delta = 200 \mu$eV, normal density of states at the Fermi energy\cite{Kittel:2004vf,Court:2008vi} $N(0)=1.4477\times10^{47}\, \rm{m}^{-3}\, \rm{J}^{-1}$ and island volume $V=0.0014\mu$m$^3$, consistent with experiments on aluminium based JJAs~\cite{Bylander:2007gb,Bylander:2005gr}.  Throughout this discussion, we consider a JJA of length $N=50$ with a ratio of junction capacitance $C_J=0.5$fF to ground capacitance $C_G=20$aF that gives a soliton length~\cite{Haviland:1996wz,Delsing:1992tq} $\Lambda = \sqrt{C_J/C_G}=5$.  In this regime the array can be considered `long' although correlated transport effects are still important~\cite{Cole:2014fk,Walker:2013wp}.

The `smoking gun' of parity effects in superconducting SETs is the observation of current plateaus at low temperature for odd charge states, while the current is suppressed completely for even charge states~\cite{Tuominen:1992vy,Amar:1994wh}.  We begin by considering the equivalent experiment for a JJA.  Fig.~\ref{fig:paritywithnodisorder} shows the I-V characteristics for a JJA as a function of effective electron temperature.  For $T_e < 170$ mK, we see only sporadic (or no) conduction as the system is too easily trapped in meta-stable states.  As $T_e$ increases, the characteristic suppression of current due to the parity effect manifests as oscillations in the current at low bias.  These oscillations stem from the interplay between the parity dependent tunnelling rates and the voltage dependent filling factors on each island.  When $V\gtrsim22$mV we see the step in current associated with the breaking of Cooper-pairs at every junction, ie.\ when $V\ge N\times2\Delta = 20$mV.  

Interestingly, the parity oscillations vanish for $V>13$mV although the magnitude of the current is still approximately constant below $T^{*}$ but increases rapidly for $T_e>T^{*}$.  At first glance this would appear to be a transition associated with $V=N\times\Delta$ however it actually depends on the interplay of charging energy and parity effects.  At low voltages, dipole states can form which are stable for certain combinations of voltage parity and it is these states which block the flow of current.  Above a certain voltage, these metastable states can dissociate via interactions with neighbouring charges - leading to more robust conduction.  One can think of this in terms of a phase-space argument where the number of available states grows with increasing voltage, therefore allowing the system to avoid getting trapped in local minima. 

\begin{figure} [t!]
\centering{\includegraphics[width=1.0\columnwidth]{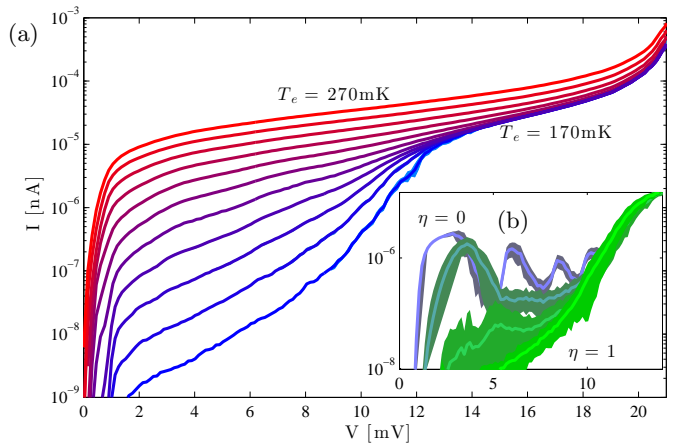}}
\rput[tl](-4.5,6){(a)}
\rput[tl](2,3.5){(b)}
\caption{(a) Current as a function of voltage and electron temperature for maximal disorder.  Below the parity temperature, the current is strongly suppressed at low voltages.  The oscillations in the current due to parity effects are also lost due to disorder. (b) The current at $T_e = 170$ mK as a function of disorder strength.  In this subfigure we ensemble averaging over 50 \emph{different disorder realisations}, the variance of which is show via pale shading.  Although there is considerable variation with disorder realisation, qualitatively we see that using $\eta=1$ insures numerical convergence for a given realisation.\label{fig:increasingdisorder}}
\end{figure}

Experimentally, background charges within the device and substrate lead to random offset charges~\cite{ZorinAB:1996uw,Johansson:2001tc,Maisi:2009eq,Zimmerman:2009dv,Zimmerman:2008ig,Wolf:1997wp,Pourkabirian:2014br,Vogt:2014vi}.  We can model this disorder as random offset charges on the islands of the JJA~\cite{Johansson:2001tc}, $|q_{bk}|/e \le \eta$, which we assume to be static on the time scale required to measure a single current point.  In Fig.~\ref{fig:increasingdisorder} we see the effects of increasing disorder is to suppress conduction for small bias, as well as eliminating the parity dependent oscillations as a function of voltage bias.  We see convergence of the response as a function of disorder strength (inset to Fig.~\ref{fig:increasingdisorder}), with $\eta\gtrsim0.7$ being sufficient to model `maximal disorder'~\footnote{Experimentally, one would assume that maximal disorder is reached at $\eta=0.5$ as tunnelling of single charges can eliminate any greater disorder during cool down.  However, here we are finding the degree of disorder required for \emph{numerical} convergence in the simulation, which reflects the ability of the system to escape from local metastable states given the processes included in the simulation.}.  As well as the issues of rare-events at low bias discussed for the non-disordered case, in the inset to Fig.~\ref{fig:increasingdisorder} we compute the variance in the current as a function of 50 different disorder realisations.  We see the same qualitative suppression of the parity oscillations with increasing disorder, even taking into account the variance due to disorder.  For all subsequent simulations (as well as the main Fig.~\ref{fig:increasingdisorder}) we take the disorder realisation to be constant (with $\eta = 1$) for a given I-V curve.  This corresponds to the case where the charge disorder is maximal but stable over the entire experimental timescale of interest.

\begin{figure} [t!]
\centering{\includegraphics[width=1.0\columnwidth]{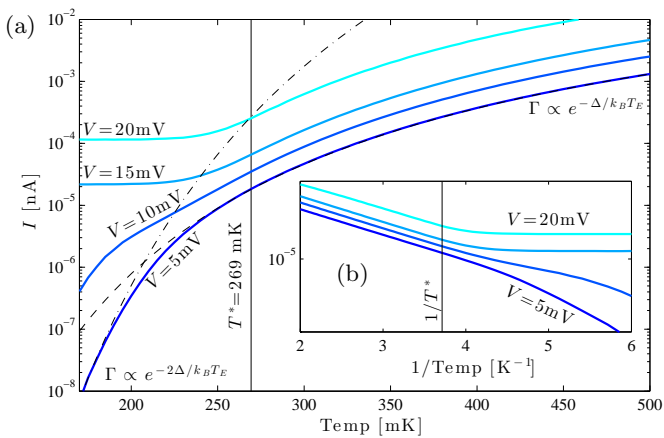}}
\rput[tl](-4.5,6){(a)}
\rput[tl](-0.1,2.7){(b)}
\caption{(a) Current plotted as a function of $T_e$ for various values of $V$.  Here we see at low voltages a characteristic crossover from $2\Delta/k_B T_e$ (dashed-dotted line) to $\Delta/k_B T_e$ (dashed line) scaling, indicating parity limited conduction.  At higher voltages, the increased state space means that the current is no longer limited by the even-even rate but is instead dominated by the odd parity processes, showing an initial constant scaling with temperature.  Irrespective of voltage (while still in the sub-gap region), at temperatures above $T^{*}$ the current scales with the parity independent rate.  (b) The inset shows the same data plotted as a function of inverse temperature, which clearly shows the two regimes.\label{fig:tempscaling}}
\end{figure}

Although the characteristic parity oscillations are not visible in the presence of maximal disorder, the overall scaling behaviour of the current is still a strong function of the parity dependent rates.  This scaling behaviour at low bias can be simply understood in terms of the scaling of the NEQ rates shown in Fig.~\ref{fig:rates}.  To observe the scaling of the tunnelling rate above and below $T^{*}$ ie.\ $\propto \exp(-\Delta/k_BT_e)$ and $\propto \exp(-2\Delta/k_BT_e)$ respectively; in Fig.\ref{fig:tempscaling} we plot the current at fixed voltage.  In the low bias regime ($V=5$mV) we see clear evidence of the cross-over from $2\Delta/kT_e$ to $\Delta/kT_e$ scaling.  This regime is where charge-charge correlation effects are strongest due to the low filling of the array~\cite{Ho:2010ft,Cole:2014fk,Walker:2013wp} which is also the regime that should show parity effects most clearly.  

In the low bias regime, we interpret this cross-over as single isolated charges moving through the array largely independently.  However, due to the background charge, the effective parity of the islands they encounter varies and therefore the slowest rate (the even-even rate) is the limiting factor.  The scaling behaviour therefore mimics that of the even-even rate below and above the parity temperature.  At higher bias, the transition from constant to $\Delta/kT_e$ scaling illustrates that the dominant tunnelling events are associated with odd-odd and odd-even processes.  In this case the higher applied bias pushes charges closer together which increases the available state space and the system can escape from even parity states which would otherwise limit conduction.  Such a transition from constant to $\Delta/kT_e$ scaling was recently seen~\cite{Zimmer:2013dg} in zero-bias conductance experiments on flux tuneable Josephson junction arrays - which suggests a thermal quasiparticle origin for the temperature dependence of the conductance.

\begin{figure} [t!]
\centering{\includegraphics[width=1.0\columnwidth]{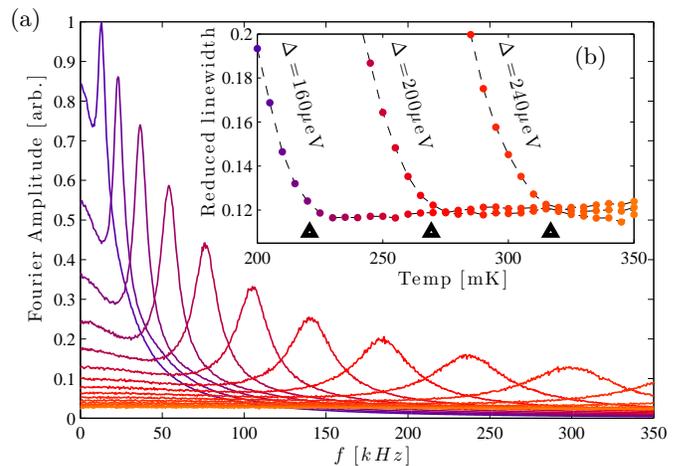}}
\rput[tl](-4.5,6.5){(a)}
\rput[tl](3,6){(b)}
\caption{(a) Fourier transform of the charge-charge correlation function as a function of temperature.  As the temperature is increased, the mean transition rate and therefore the current also increases.  As the temperature is increased, correlated conduction sets in soon after conduction itself begins.  The ratio of correlation frequency to current $f_{peak}/I=e$ is consistent with the charge carriers being single electrons.  (b)  The width of the correlation peaks normalised by the position (the reduced line width) shows a particularly clear signature of the parity effect.  As a function of temperature we see the reduced line width plateau above the parity temperature (indicated with triangles for different values of the superconducting gap).~\label{fig:correlatedtransport}}
\end{figure}

Recent experiments have demonstrated counting of individual electrons within a JJA~\cite{Bylander:2005gr,Bylander:2007gb} - which raises the question of whether signatures of parity can be seen in the charge-charge correlations during transport.  Even in conventional (normal) conducting JJAs, there is strong interplay between applied voltage and correlated transport (through the average charge density).  To focus on the role of parity, we compute the charge-charge correlation function~\cite{Bakhvalov:1989td,Cole:2014fk} on site number 15 of the array for a fixed applied voltage $V=2$mV, sampled with a bandwidth of 5 MHz.  As a function of temperature (at fixed voltage) we see three distinct regions, see Fig.~\ref{fig:correlatedtransport}.  

At very low temperatures and currents, we see no correlation as the transport is too slow on the time scale of the simulations.  As the temperature and therefore current increases, strongly correlated transport sets in with the correlation peak frequency scaling linearly with current according to $I = e f_{\rm{peak}}$, reflecting the fact that the charge carriers are single electrons.  Ultimately the amplitude of the correlation peak reduces due to increasing charge noise at high currents - showing a surprising similarity as a function of temperature to the `washout' seen at high voltage bias~\cite{Cole:2014fk}.

More subtly, the role of the parity effect can be seen in the `reduced line width' of the response, ie.\ peak width/peak position, as a function of temperature (inset to Fig.\ref{fig:correlatedtransport}).  Once correlated transport sets in, the peak width reduces as a function of increasing temperature until it reaches a constant value when $T_e\approx T^{*}$, which is largely independent of the value of the superconducting gap.  We can ascribe this step behaviour to the additional noise in the correlation signal due to the parity effect, which effectively vanishes once $T_e$ reaches $T^{*}$.

\section{Conclusion}
Josephson junction arrays provide a tantalising playground for studying many-body effects as they are a controllable, artificial system which is truly one-dimension and yet displays correlation electron effects.  It is therefore supremely disappointing that experimental results to date can only be explained qualitatively at best.  An important contributor to this situation is the difficulty in (experimentally) filtering out or (theoretically) accounting for quasiparticle effects.  We have shown how to model the transition from many to single quasiparticle excitations in multi-junction circuits, in particular we considered the quasiparticle parity in a consistent way.  Even when considering the limit of strongly disordered offset charges, the crossover to the parity regime can be observed in both the current-voltage characteristics and the charge-charge correlation function.

\section{Acknowledgements}
This work was supported by the Victorian Partnership for Advanced Computing (VPAC).  TD is supported by the ARC Centre of Excellence for Engineered Quantum Systems, CE110001013.  The authors would like to acknowledge useful discussions with K. Walker, N. Vogt, D. Golubev and G. Sch\"on.
	
\bibliographystyle{unsrt}
\bibliography{NEQPJJA}

\end{document}